\documentclass[letterpaper, conference, 10pt]{IEEEtran}

\usepackage{etex}

\usepackage{ifpdf}
\ifpdf
\usepackage[draft,pdfborder={0 0 0}]{hyperref}
\fi

\usepackage{algorithm}
\usepackage{algorithmicx}
\usepackage{algpseudocode}

\usepackage[cmex10]{amsmath}
\usepackage{amssymb}
\usepackage{nicefrac}

\usepackage{bm}
\usepackage[varg]{txfonts}
\let\mathbb=\varmathbb
\DeclareSymbolFont{letters}{OML}{ztmcm}{m}{it}

\usepackage[font=footnotesize]{subfig}

\usepackage{tabularx}
\usepackage{multirow}
\usepackage{dcolumn}
\usepackage{booktabs}

\usepackage{fixltx2e}

\usepackage{tikz}
\usepackage{pgfplots}
\usetikzlibrary{plotmarks,shapes,positioning,arrows,decorations.markings,fit,calc,patterns}

\input{figures/register.tikz} 

\usepackage{glossaries}
\makeglossaries

\usepackage{balance}

\title{A Multi-Gbps Unrolled Hardware List Decoder\\ for a Systematic Polar Code}

\author{\IEEEauthorblockN{Pascal Giard\IEEEauthorrefmark{1}\IEEEauthorrefmark{2}, %
Alexios Balatsoukas-Stimming\IEEEauthorrefmark{2}, Thomas Christoph M{\"u}ller\IEEEauthorrefmark{2}, \\%
Andreas Burg\IEEEauthorrefmark{2}, Claude Thibeault\IEEEauthorrefmark{3}, and %
Warren J. Gross\IEEEauthorrefmark{1}}
  \IEEEauthorblockA{\IEEEauthorrefmark{1}Department of Electrical and Computer Engineering, McGill University, Montr\'eal, Qu\'ebec, Canada.\\Email: pascal.giard@mail.mcgill.ca, warren.gross@mcgill.ca}%
  \IEEEauthorblockA{\IEEEauthorrefmark{2}Telecommunications Circuits Laboratory, \'Ecole polytechnique f\'ed\'erale de Lausanne, Lausanne, Switzerland.\\Email: \{pascal.giard,alexios.balatsoukas,christoph.mueller,andreas.burg\}@epfl.ch}%
  \IEEEauthorblockA{\IEEEauthorrefmark{3}Department of Electrical Engineering, \'Ecole de technologie sup\'erieure, Montr\'eal, Qu\'ebec, Canada.\\Email: claude.thibeault@etsmtl.ca}
}

\usepackage[defernumbers=true,backend=bibtex,sorting=none,style=ieee]{biblatex}
\begin{document}

\newacronym{crc}{CRC}{cyclic redundancy check}
\newacronym{fer}{FER}{frame-error rate}
\newacronym{ber}{BER}{bit-error rate}
\newacronym{ldpc}{LDPC}{low-density parity-check}
\newacronym{sc}{SC}{successive-cancellation}
\newacronym{bpsk}{BPSK}{binary phase-shift keying}
\newacronym{awgn}{AWGN}{additive white Gaussian noise}
\newacronym{par}{PAR}{place-and-route}
\newacronym[plural=LLRs,firstplural=log-likelihood ratios (LLRs)]{llr}{LLR}{log-likelihood-ratio}
\newacronym{scl}{SCL}{successive-cancellation list}
\newacronym{spc}{SPC}{single-parity check}

\maketitle

\begin{abstract}
Polar codes are a new class of block codes with an explicit construction that provably achieve the capacity of various communications channels, even with the low-complexity successive-cancellation (SC) decoding algorithm. Yet, the more complex successive-cancellation list (SCL) decoding algorithm is gathering more attention lately as it significantly improves the error-correction performance of short- to moderate-length polar codes, especially when they are concatenated with a cyclic redundancy check code. However, as SCL decoding explores several decoding paths, existing hardware implementations tend to be significantly slower than SC-based decoders. In this paper, we show how the unrolling technique, which has already been used in the context of SC decoding, can be adapted to SCL decoding yielding a multi-Gbps SCL-based polar decoder with an error-correction performance that is competitive when compared to an LDPC code of similar length and rate. Post-place-and-route ASIC results for 28~nm CMOS are provided showing that this decoder can sustain a throughput greater than 10 Gbps at 468~MHz with an energy efficiency of 7.25~pJ/bit.
\end{abstract}

\section{Introduction}
Polar codes were recently selected for the next-generation mobile communications standard that is currently under development by the 3GPP due to their excellent error-correction performance at short to moderate blocklengths under SCL decoding~\cite[p. 123]{3GPPRANPolar}. Unfortunately, most SCL decoder implementations in the literature still suffer from low throughput and high decoding latency~\cite{Balatsoukas-Stimming_TSP_2015,Yuan2015,Lin2016,Xiong2016}. Several algorithmic and architectural improvements have been proposed in order to remedy this situation. For example, multi-bit SCL decoding~\cite{Yuan2015} can significantly reduce the decoding latency of SCL decoding, but at a very large cost in terms of the required hardware resources. A similar approach, which groups multiple bits into symbols and transforms the SCL decoder to a symbol-based SCL decoder was presented in~\cite{Xiong2016b} and is shown to offer similar decoding throughput improvements compared to standard multi-bit SCL decoding, but with lower decoding complexity. A different approach was taken in~\cite{Sarkis_JSAC_2016} where the proposed Fast-SSC-List decoding algorithm employs specialized decoding units for smaller sub-codes of the polar code in order to reduce the decoding latency.

Unrolled decoders are known for their tremendous throughput \cite{Schlafer2013,Giard_IET_2015,Balatsoukas-Stimming2015,Giard_TCASI_2016}. They offer at least one order of magnitude improvement in throughput with respect to standard decoders at the cost of larger area requirements. While this unrolling technique has been applied to SC-based polar decoders before~\cite{Giard_IET_2015,Giard_TCASI_2016}, it has not yet been applied to a hardware \gls{scl}-based decoder. Applying the technique to the original \gls{scl} decoding algorithm \cite{Tal2015} would result in a hardware implementation with very high area complexity. Thus, in this paper, we propose an unrolled hardware implementation of the Fast-SSC-List decoding algorithm~\cite{Sarkis_JSAC_2016}. We show that, for a $(512, 427)$ systematic polar code, the throughput is an order of magnitude higher than the state of the art, while the error-correction performance is better than that of the $(576, 480)$ \gls{ldpc} code from the IEEE 802.16e standard \cite{WiMAX}.

\textit{Outline:} The remainder of this paper starts with Section~\ref{sec:bg} by providing the necessary background, consisting of a brief review of polar codes and an introduction to \gls{scl}-based decoding algorithms. Moreover, we present a comparison of the error-correction performance of an \gls{scl}-decoded polar code against that of an \gls{ldpc} code from the IEEE 802.16e standard. Section~\ref{sec:bg} also briefly reviews the Fast-SSC-List decoding algorithm. Section~\ref{sec:arch} describes our adaptation of the fully-unrolled and pipelined hardware architecture to \gls{scl} decoding. Section~\ref{sec:impl} discusses implementation details and provides post-\gls{par} ASIC area, timing, and power results for the 28 nm UTBB-FD-SOI CMOS technology from ST Microelectronics. A comparison against the state-of-the-art \gls{scl}-based decoder implementations from the literature is also carried out in Section~\ref{sec:impl}. Finally, Section~\ref{sec:conclusion} concludes this paper.

\section{Background}\label{sec:bg}
\subsection{Polar Codes}
In his original work, Ar{\i}kan used a linear transformation of a vector of bits that can be shown to lead to a polarization phenomenon, meaning that some of these bits experience almost noiseleses transmission channels while the remaining bits experience almost completely noisy transmission channels. Polar codes exploit this polarization phenomenon to achieve the symmetric capacity of memoryless channels as the code length goes to infinity. More specifically, to construct an ($N$, $k$) polar code, the $N-k$ least reliable bits (i.e., the bits that experience the $N-k$ worst transmission channels), called the frozen bits, are set to zero and the remaining $k$ bits are used to carry actual information.

Polar codes provably achieve capacity when decoded using the low-complexity \gls{sc} algorithm~\cite{Arikan2009}. However, with \gls{sc} decoding, the error-correction performance of polar codes at short to moderate length is in general worse than the error-correction performance of other modern channel codes. It was shown that decoding polar codes using an \gls{scl}-based decoding algorithm significantly improves the situation~\cite{Tal2015}, especially when concatenating the polar code with a \gls{crc}~\cite{Niu2012,Tal2015}.

It was shown in \cite{Arikan2011} that polar codes can be encoded and decoded systematically, leading to an improved \gls{ber} without affecting the \gls{fer}. In this work, systematic polar codes are used.

\subsection{Successive-Cancellation List Decoding}

The \gls{sc} decoding algorithm is a greedy algorithm: it uses the channel output $\bm{y}_0^{N-1}$ and the previous bit estimates $\hat{\bm u}_0^{i-1}$ to estimate the value of bit $\hat{u}_i$. Therefore, as soon as an error occurs a frame will inevitably be in error as past decisions are never revisited. \gls{scl}-based decoding algorithms for polar codes are also greedy in the sense that they sequentially build the most likely codewords. However, at each step, instead of considering only the most likely bit value, both possible bit values---0 and 1---are considered. Thus, as the decoding proceeds a constrained list of up to $L$ potential candidate codewords is built, and a reliability metric is calculated for each path along the way. At the very end, the most likely codeword among the candidates in the list is selected. In the case of the CRC-aided \gls{scl} (CA-SCL) decoding algorithm the polar code is concatenated with a \gls{crc} and when decoding ends the \gls{crc} is calculated for all $L$ candidate codewords. The most likely codeword with a calculated \gls{crc} that matches the expected CRC is selected as the estimated codeword. If none of the \glspl{crc} matches, the codeword with the best reliability is selected.

\begin{figure}[t]
  \centering
  \usetikzlibrary{plotmarks}

\definecolor{lbluecb}{RGB}{86, 180, 233}
\definecolor{orangecb}{RGB}{213, 94, 0}

\begin{tikzpicture}

  \pgfplotsset{
    grid style = {
      dash pattern = on 0.05mm off 1mm,
      line cap = round,
      black,
      line width = 0.5pt
    },
    label style = {font=\fontsize{10pt}{7.2}\selectfont},
    tick label style = {font=\fontsize{9pt}{7.2}\selectfont}
  }

 \begin{semilogyaxis}[%
    xlabel=$E_b/N_0$ (dB),xtick={2,2.5,...,5.0},%
    xlabel style={yshift=0.6em},%
    minor x tick num={1},
    xmin=3,xmax=5,%
    ylabel=FER, ylabel style={yshift=-1.05em},%
    width=0.54\columnwidth, height=7cm, grid=major,%
    legend style={
      anchor={center},
      cells={anchor=west},
      column sep=2mm,
      font=\fontsize{8pt}{7.2}\selectfont,
      mark size=3.0pt
    },
    legend columns=4,
    legend to name=perf-legend,
    mark size=3.0pt]
    
    \addlegendimage{empty legend}
    \addlegendentry[anchor=east]{\textbf{List:}}

    \addplot[very thick,color=black] table[x=ebn0_db,y=FER] {data/512.427.s0.5.list.L2.C0.float.csv};
    \addlegendentry{$L = 2$}

    \addlegendimage{empty legend}
    \addlegendentry{}

    \addlegendimage{empty legend}
    \addlegendentry{}

    \addlegendimage{empty legend}
    \addlegendentry[anchor=east]{\textbf{List-CRC:}}

    \addplot[very thick,color=orangecb, mark=x] table[x=ebn0_db,y=FER] {data/512.427.s0.5.list.L2.C8.float.csv};
    \addlegendentry{$L = 2$}

    \addlegendimage{empty legend}
    \addlegendentry{}

    \addlegendimage{empty legend}
    \addlegendentry{}

    \addlegendimage{empty legend}
    \addlegendentry[anchor=east]{\textbf{LDPC:}}

    \addplot[very thick,color=lbluecb, mark=o, dashed, mark options={solid}] table[x=ebn0_db,y=FER] {data/wimax_576_480_SCMS_05It.csv};
    \addlegendentry{$I=5$}

    \addplot[very thick,color=lbluecb, mark=diamond, dashed, mark options={solid}] table[x=ebn0_db,y=FER] {data/wimax_576_480_SCMS_10It.csv};
    \addlegendentry{$I=10$}

    \addplot[very thick,color=lbluecb, mark=triangle, dashed, mark options={solid}] table[x=ebn0_db,y=FER] {data/wimax_576_480_SCMS_15It.csv};
    \addlegendentry{$I=15$}

  \end{semilogyaxis}
\end{tikzpicture}\hspace{-7pt}
\begin{tikzpicture}
  \pgfplotsset{
    grid style = {
      dash pattern = on 0.05mm off 1mm,
      line cap = round,
      black,
      line width = 0.5pt
    },
    label style = {font=\fontsize{10pt}{7.2}\selectfont},
    tick label style = {font=\fontsize{9pt}{7.2}\selectfont}
  }

 \begin{semilogyaxis}[%
    xlabel=$E_b/N_0$ (dB),xtick={2,2.5,...,5.0},%
    xlabel style={yshift=0.6em},%
    minor x tick num={1},
    xmin=3,xmax=5,%
    ylabel=BER, ylabel style={yshift=-1.05em},%
    width=0.54\columnwidth, height=7cm, grid=major,%
    mark size=3.0pt]
    
    \addplot[very thick,color=black] table[x=ebn0_db,y=BER] {data/512.427.s0.5.list.L2.C0.float.csv};

    \addplot[very thick,color=orangecb, mark=x] table[x=ebn0_db,y=BER] {data/512.427.s0.5.list.L2.C8.float.csv};

    \addplot[very thick,color=lbluecb, mark=o, dashed, mark options={solid}] table[x=ebn0_db,y=BER] {data/wimax_576_480_SCMS_05It.csv};

    \addplot[very thick,color=lbluecb, mark=diamond, dashed, mark options={solid}] table[x=ebn0_db,y=BER] {data/wimax_576_480_SCMS_10It.csv};

    \addplot[very thick,color=lbluecb, mark=triangle, dashed, mark options={solid}] table[x=ebn0_db,y=BER] {data/wimax_576_480_SCMS_15It.csv};

  \end{semilogyaxis}
\end{tikzpicture}
\\
\ref{perf-legend}
  \caption{Error-correction performance of SCL and CA-SCL decoding of a $(512, 427)$ systematic polar code versus that of the $(576, 480)$ IEEE 802.16e LDPC code. $L$ is the maximum number of candidate codewords for the SCL decoding algorithm and $I$ is the maximum number of iterations for the self-corrected min-sum algorithm.}
  \label{fig:perf-cmp}
\end{figure}
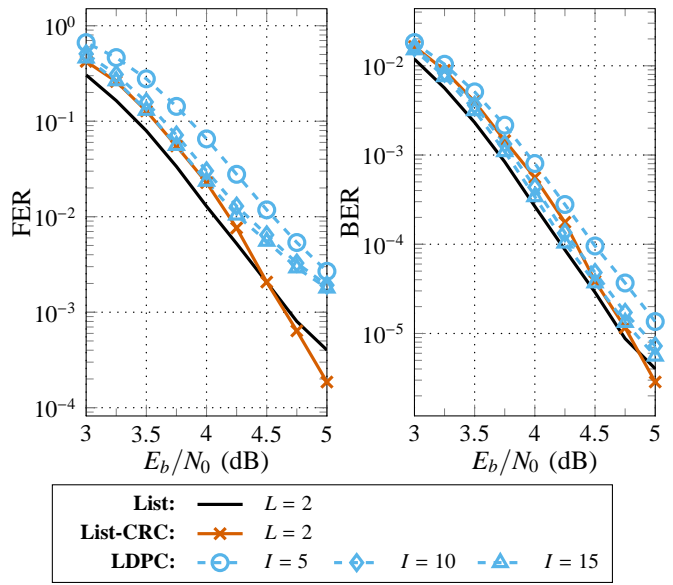

Fig.~\ref{fig:perf-cmp} shows the error-correction performance of a $(512, 427)$ systematic polar code decoded using the \gls{scl} and the CA-\gls{scl} algorithm with $L=2$. An 8-bit CRC is used for the CA-\gls{scl} decoding algorithm. The performance is simulated for random codewords, using a \gls{bpsk} modulation over an \gls{awgn} channel. Both \gls{fer} and \gls{ber} of the $(576, 480)$ \gls{ldpc} code from the IEEE 802.16e standard \cite{WiMAX} are included for comparison. The \gls{ldpc} code is decoded with a layered schedule using the self-corrected min-sum algorithm~\cite{Savin2008}. It can be seen that the chosen polar code compares favorably against an \gls{ldpc} code of similar length and rate even with a list size as small as $L=2$. It should also be noted that, in this particular scenario, for a targeted \gls{fer} of $10^{-3}$, it is beneficial to \textit{not} concatenate the polar code with a \gls{crc}. This observation was also made in \cite{Balatsoukas-Stimming_TSP_2015} i.e., in some cases, it is more beneficial---error-rate wise---to not concatenate a polar code with a \gls{crc}.

Unfortunately, \gls{scl} decoding involves the exploration of multiple decoding paths simultaneously as well as a costly path metric sorting step. Thus, hardware implementations of \gls{scl} decoding are typically much slower than state-of-the-art \gls{sc}-based decoders. As mentioned in the introduction, multiple algorithms employing multi-bit decisions have been proposed \cite{Yuan2015,Sarkis_JSAC_2016,Lin2016,Xiong2016b,Xiong2016} to significantly reduce the decoding latency and increase the decoding throughput of \gls{scl} decoding. The hardware implementation presented in this paper is based on the Fast-SSC-List decoding algorithm proposed in \cite{Sarkis_JSAC_2016}.

\begin{figure*}[th]
  \centering
  \begin{tikzpicture}[font=\footnotesize,inner sep=1pt, minimum width=1.2em]

  \definecolor{deepgreen}{RGB}{8, 130, 25}

\tikzset{
branch/.style={fill,shape=circle,minimum size=3pt,inner sep=0pt},
block/.style={draw, rectangle, minimum height=2em},
spc/.style={draw, rectangle, minimum height=2em},
rep/.style={draw, rectangle, minimum height=2em},
comb/.style={draw, rectangle, minimum height=3.5em},
best/.style={draw, rectangle, minimum height=10em},
}

\node (ac) at (0.2,0) {$\alpha_0^{7}$};

\node[shape=reg] at ($(ac)+(0.775,-0.195)$) (REGAC1) {$\alpha_c$};

\node[shape=reg] at ($(REGAC1)+(1.8,0)$) (REGG01) {$\alpha_2$};
\node[shape=reg] at ($(REGAC1)+(1.8,-2.09)$) (REGG11) {$\alpha_3$};
\node[block] at ($(REGAC1)+(0.9,-3.31)$) (F1) {$F$};
\node[block] at ($(REGAC1)+(0.9,0.19)$) (G01) {$G_0$};
\node[block] at ($(REGAC1)+(0.9,-1.9)$) (G11) {$G_1$};

\node[shape=reg] at ($(F1)+(0.9,-0.189)$) (REGF) {$\alpha_1$};

\node[shape=reg] at ($(REGG01)+(1.9,0)$) (REGG02) {$\alpha_2$};
\node[shape=reg] at ($(REGG11)+(1.9,0)$) (REGG12) {$\alpha_3$};
\node[block,fill=deepgreen] at ($(REGF)+(0.9,0.19)$) (Rep1) {$Rep$};
\node[shape=reg] at ($(Rep1)+(1.0,0.0)$) (REGPM0) {\tiny$PM^0$};
\node[shape=reg] at ($(Rep1)+(1.0,-0.95)$) (REGL0) {$\ell^0$};
\node (pm0) at ($(REGPM0.Q)+(0.5,0)$) {$PM^0$};
\node (l0) at ($(REGL0.Q)+(0.45,0)$) {$\ell^0$};

\node[spc,fill=orange] at ($(REGG02)+(1.6,0)$) (SPC0) {\shortstack{SPC\\{\tiny 0}}};
\node[spc,fill=orange] at ($(REGG12)+(1.6,0)$) (SPC1) {\shortstack{SPC\\{\tiny 1}}};
\node (spc0pmin) at ($(SPC0.west)-(0.6,0.2)$) {$PM_0^0$};
\node (spc1pmin) at ($(SPC1.west)-(0.6,0.2)$) {$PM_1^0$};
\node[shape=reg] at ($(SPC0)+(2.0,0.0)$) (REGPM1) {\tiny$PM^1$};
\node[shape=reg] at ($(SPC0)+(2.0,-.95)$) (REGL1) {$\ell^1$};
\node[shape=reg] at ($(SPC1)+(2.0,0.0)$) (REGPM2) {\tiny$PM^2$};
\node[shape=reg] at ($(SPC1)+(2.0,-.95)$) (REGL2) {$\ell^2$};
\node[block] at ($(REGL1)-(1.0,-0.2)$) (Concat0) {$\&_0$};
\node[block] at ($(REGL2)-(1.0,-0.2)$) (Concat1) {$\&_1$};
\node (concat0lin) at ($(Concat0.west)-(0.6,0.2)$) {$\ell_0^0$};
\node (concat1lin) at ($(Concat1.west)-(0.6,0.2)$) {$\ell_1^0$};

\node[best] at ($(REGL1)+(0.9,-0.40)$) (Best1) {\rotatebox{90}{$L$-Best Candidates}};
\node[shape=reg] at ($(REGL1)+(1.8,0)$) (REGPM3) {\tiny$PM^3$};
\node[shape=reg] at ($(REGL1)+(1.8,-0.95)$) (REGL3) {$\ell^3$};

\node[comb] at ($(REGL3)+(0.9,0.2)$) (Comb1) {\rotatebox{90}{$Combine$}};
\node[shape=reg] at ($(REGPM3)+(1.8,0)$) (REGPM4) {\tiny$PM^4$};
\node[shape=reg] at ($(REGL3)+(1.8,0)$) (REGL4) {$\ell^4$};

\node[comb, minimum height=6em] at ($(REGPM4)+(0.9,-0.3)$) (Best2) {\rotatebox{90}{Best Candidate}};
\node[shape=reg] at ($(Best2)+(0.8,0)$) (REGBC) {$\beta_c$};

\node (bc) at ($(REGBC.Q)+(0.7,0)$) {$\beta_0^{7}$};


\draw[-] (ac.east) -- (REGAC1.D);

\draw[dotted] ($(REGAC1.Q)+(-0.25,0.5)$) -- ($(REGAC1.Q)+(-0.25,0.2)$) ($(REGAC1.Q)+(-0.25,-0.6)$) -- ($(REGAC1.Q)+(-0.25,-5.2)$);

\draw[-] (REGAC1.Q) -- (G01) -- (REGG01.D);
\node (ac1branch) at ($(REGAC1.Q) + (0.1,0)$) {};
\draw[-] (ac1branch) node[branch] {};
\node (ac1branch) at ($(REGAC1.Q) + (0.1,0)$) {};
\draw[-] (ac1branch) |- (ac1branch |- G11) node[branch] {} -- (G11) -- (REGG11.D);
\draw[-] (ac1branch |- G11) |- (ac1branch |- F1) |- (F1) -- (REGF.D);

\draw[dotted] ($(REGG01.Q)+(-0.25,0.5)$) -- ($(REGG01.Q)+(-0.25,0.2)$) ($(REGG01.Q)+(-0.25,-0.6)$) -- ($(REGG11.Q)+(-0.25,0.2)$) ($(REGG11.Q)+(-0.25,-0.6)$) -- ($(REGF.Q)+(-0.25,0.2)$) ($(REGF.Q)+(-0.25,-0.6)$) -- ($(REGG01.Q)+(-0.25,-5.2)$);

\draw[-] (REGG01.Q) -- (REGG02.D);
\draw[-] (REGG11.Q) -- (REGG12.D);
\draw[-] (REGF.Q) -- (Rep1);
\draw[-] (Rep1.east |- REGPM0.D) -- (REGPM0.D);
\draw[-] (Rep1) -- ($(Rep1.east)+(0.2,0)$) |- (REGL0.D);

\draw[dotted] ($(REGG02.Q)+(-0.25,0.5)$) -- ($(REGG02.Q)+(-0.25,0.2)$) ($(REGG02.Q)+(-0.25,-0.6)$) -- ($(REGG12.Q)+(-0.25,0.2)$) ($(REGG12.Q)+(-0.25,-0.6)$) -- ($(REGPM0.Q)+(-0.25,0.2)$) ($(REGPM0.Q)+(-0.25,-0.6)$) -- ($(REGL0.Q)+(-0.25,0.2)$) ($(REGL0.Q)+(-0.25,-0.6)$) -- ($(REGG02.Q)+(-0.25,-5.2)$);

\draw[-] (REGG02.Q) -- (SPC0.west |- REGG02.Q);
\draw[-] (spc0pmin.east) -- (SPC0.west |- spc0pmin);
\draw[-] (REGG12.Q) -- (SPC1.west |- REGG12.Q);
\draw[-] (spc1pmin.east) -- (SPC1.west |- spc1pmin);
\draw[-] (SPC0.east |- REGPM1.D) -- (REGPM1.D);
\draw[-] (SPC1.east |- REGPM2.D) -- (REGPM2.D);
\draw[-] (SPC0) -- ($(SPC0.east)+(0.2,0)$) |- (Concat0) -- (REGL1.D);
\draw[-] (SPC1) -- ($(SPC1.east)+(0.2,0)$) |- (Concat1) -- (REGL2.D);
\draw[-] (concat0lin.east) -- (Concat0.west |- concat0lin);
\draw[-] (concat1lin.east) -- (Concat1.west |- concat1lin);
\draw[-] (REGPM0.Q) -- (pm0.west);
\draw[-] (REGL0.Q) -- (l0.west);

\draw[dotted] ($(REGPM1.Q)+(-0.25,0.5)$) -- ($(REGPM1.Q)+(-0.25,0.2)$) ($(REGPM1.Q)+(-0.25,-0.6)$) -- ($(REGL1.Q)+(-0.25,0.2)$) ($(REGL1.Q)+(-0.25,-0.6)$) -- ($(REGPM2.Q)+(-0.25,0.2)$) ($(REGPM2.Q)+(-0.25,-0.6)$) -- ($(REGL2.Q)+(-0.25,0.2)$) ($(REGL2.Q)+(-0.25,-0.6)$) -- ($(REGPM1.Q)+(-0.25,-5.2)$);

\draw[-] (REGPM1.Q) -- (Best1.west |- REGPM1.Q);
\draw[-] (REGL1.Q) -- (Best1.west |- REGL1.Q);
\draw[-] (REGPM2.Q) -- (Best1.west |- REGPM2.Q);
\draw[-] (REGL2.Q) -- (Best1.west |- REGL2.Q);
\draw[-] (REGPM3.D) -- (Best1.east |- REGPM3.D);
\draw[-] (REGL3.D) -- (Best1.east |- REGL3.Q);

\node (REGPM1anc0) at ($(REGPM1.Q)+(-0.25,0.5)$) {};
\node (REGPM1anc1) at ($(REGPM1.Q)+(-0.25,-5.2)$) {};
\draw[dotted] (REGPM1anc0 -| REGPM3) -- ($(REGPM3.Q)+(-0.25,0.2)$) ($(REGPM3.Q)+(-0.25,-0.6)$) -- ($(REGL3.Q)+(-0.25,0.2)$) ($(REGL3.Q)+(-0.25,-0.6)$) -- (REGPM1anc1 -| REGPM3);

\draw[-] (REGPM3.Q) -- (REGPM4.D);
\draw[-] (REGL3.Q) -- (Comb1.west |- REGL3.Q) (Comb1.east |- REGL4.D) -- (REGL4.D);

\draw[dotted] (REGPM1anc0 -| REGPM4) -- ($(REGPM4.Q)+(-0.25,0.2)$) ($(REGPM4.Q)+(-0.25,-0.6)$) -- ($(REGL4.Q)+(-0.25,0.2)$) ($(REGL4.Q)+(-0.25,-0.6)$) -- (REGPM1anc1 -| REGPM4);

\draw[-] (REGPM4.Q) -- (Best2.west |- REGPM4.Q);
\draw[-] (REGL4.Q) -- (Best2.west |- REGL4.Q);
\draw[-] (Best2.east |- REGBC.D) -- (REGBC.D);

\draw[dotted] (REGPM1anc0 -| REGBC) -- ($(REGBC.Q)+(-0.25,0.2)$) ($(REGBC.Q)+(-0.25,-0.6)$) -- (REGPM1anc1 -| REGBC);

\draw[-] (REGBC.Q) |- (bc.west);

\end{tikzpicture}
  \caption{Fully-unrolled deeply-pipelined Fast-SSC-List decoder for a $(8, 4)$ polar code with $L=2$. Constituent decoders for a Repetition code and for SPC codes are shown in green and orange, respectively. Clock signals omitted for clarity.}
  \label{fig:unrolled-hw}
\end{figure*}
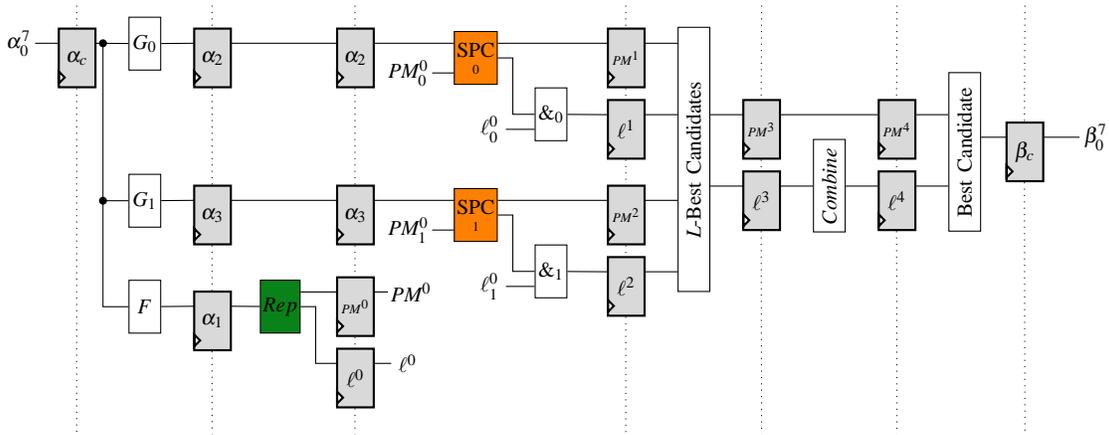

\subsection{Fast-SSC-List Decoding}
\gls{scl} decoding uses the \gls{sc} decoding algorithm, a sequential algorithm proceeding bit by bit, which effectively limits the achievable speed of hardware implementations. Recognizing that a polar code is the concatenation of smaller constituent codes, it was shown in \cite{Sarkis_JSAC_2014} that many constituent codes  could be more efficiently estimated with dedicated decoders compared to using the processing elements implementing the \gls{sc} algorithm. This led to the Fast-SSC decoding algorithm, where multiple bits are estimated simultaneously. In \cite{Sarkis_JSAC_2016}, it was proposed to adapt the Fast-SSC algorithm to \gls{scl} decoding.

Fast-SSC-List decoding provides algorithms for four different constituent codes: Rate-1, Rate-0, Repetition, and \gls{spc}. Each algorithm consists of two parts. The first part is the candidate generation, it consists of creating the $L$ most likely bit estimate vectors $\beta$. The second part consists of computing the corresponding path reliability metrics $PM$. Note that for a Rate-0 constituent code, only one path reliability metric is computed as there is only one possible candidate estimated bit vector, the all-zero vector.

While the proposed algorithms for the Rate-0 and Repetition codes are exact, approximations are used for Rate-1 and SPC codes. Thus, although it can be kept small, there is some coding loss inherent to the Fast-SSC-List decoding algorithm compared to the \gls{scl} algorithm.

\section{Unrolled and Pipelined SCL Decoder Architecture}\label{sec:arch}
As we have already mentioned, unrolled decoder architectures provide extremely high decoding speeds. In an unrolled decoder architecture, each and every operation required is instantiated in hardware so that data can flow through the decoder with minimal control. An unrolled and pipelined architecture for \gls{sc}-based polar decoding was first described in~\cite{Giard_IET_2015}, and later improved and generalized in~\cite{Giard_TCASI_2016}. In this section, we explain how the unrolled fast-SC decoder architecture of~\cite{Giard_TCASI_2016} can be extended to Fast-SSC-List decoding~\cite{Sarkis_JSAC_2016} by means of a small example.

Fig.~\ref{fig:unrolled-hw} shows an implementation example of a fully-unrolled deeply-pipelined Fast-SSC-List-based decoder for an $(8,4)$ polar code with $L=2$. The $F$ and $G$ blocks, generating the soft-input \glspl{llr} to the constituent decoders, implement the same functions as in the \gls{sc} algorithms using the min-sum approximation \cite{Leroux2013}. The $Combine$ block corresponds to one stage of a polar encoder. The ``$\&$'' blocks are bit-vector joining operators, and registers are shown in light gray, a Repetition node in green and \gls{spc} nodes in orange. 

Each register denoted $\ell$ is used to store one of the paths that survived, expressed as a partial sum. The $\ell$-registers at the ouput of a concatenation block ``$\&$'' have the surviving paths concatenated with the $L$ new bit estimate vectors $\beta$ coming out of the consituent decoders. Thus, at the output of a sorting step---denoted $L$-Best Candidate in Fig.~\ref{fig:unrolled-hw}---a register $\ell$ contains the $L$ paths that survived, also expressed as a partial sum. Thus a $Combine$ block updates the partial sum by operating on the right hand side bits of an $\ell$-register.

In this example, the $G$ functions are calculated preemptively as a Repetition node allows for only 2 possible outcomes and as the first path fork occurs there, the generated paths at the output of the Repetition node will necessarily be retained among the $L=2$ best candidates. For the general case however, this cannot be applied as the newly estimated paths (or partial sums) may be combined with different path sources before entering a $G$ function.

While not illustrated in Fig.~\ref{fig:unrolled-hw}, a figure depicting a very small polar code, as soon as \glspl{llr} $\alpha$ need to be retained past a sorting step, multiplexers have to be inserted after each sorting block. Those multiplexers allow the decoder to select the \gls{llr} values corresponding to the surviving path sources.

\section{Implementation and Results}\label{sec:impl}
For this paper, we have implemented a fully-unrolled partially-pipelined Fast-SSC-List decoder with an initiation interval $\mathcal{I}=20$ and a list size $L=2$. An initiation interval $\mathcal{I}=20$ means that a new frame is fed into the decoder every $20$ clock cycles. It also means that a new estimated codeword is available at the output of the decoder every $20$ clock cycles. The Rate-0, Repetition, and \gls{spc} nodes were constrained to a maximum length of 8, 8, and 4, respectively. The information (i.e., rate-1) nodes were not constrained and as a result, the largest one has a length of 128. The \gls{spc} node is pipelined over 2 clock cycles to shorten its longest path.

The critical path of the decoder is the sorting block---denoted ``$L$-Best Candidates'' in Fig.~\ref{fig:unrolled-hw}---; it starts from the output of a register storing a path metric, then through 2 levels of 7-bit comparators and ends into another register storing one the 2 best path metrics. The clock frequency was selected to obtain an information throughput of 10~Gbps.

\subsection{Methodology}

ASIC post-\gls{par} results are for the 28 nm UTBB-FD-SOI CMOS technology from ST Microelectronics using the RVT standard-cell library. Synopsys Design Compiler and Cadence Innovus were used for synthesis and \gls{par}, respectively. Clock gating was used in order to reduce power consumption. The post-\gls{par} netlist has been simulated for verification as well as for the generation of vectors. These latter ones have then been used to annotate the design data with toggle rates that correspond to steady state operation---i.e. a filled pipeline---in order to extract a meaningful estimation of the average power consumption of 250 random frames. In our results, the decoding latency includes the time required to load the channel \glspl{llr}, decode a frame and output the best estimated codeword.

\subsection{Impact of Quantization}

All \glspl{llr} and path metrics are expressed using the two's complement representation. The \gls{llr} value quantization is denoted as $Q_i.Q_c.Q_f$, where $Q_c$ is the total number of bits used to store a channel \gls{llr}, $Q_i$ is the total the number of bits used to store internal LLRs and $Q_f$ is the number of fractional bits in both types of \glspl{llr}. The number of bits used to represent a path metric is $Q_i+1$, and path metrics are normalized after each sorting step in order to avoid overflows.

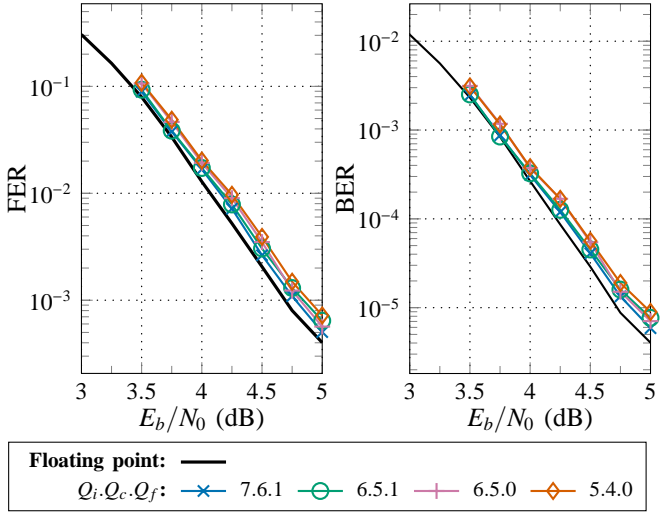
\begin{figure}[t]
  \centering
  \usetikzlibrary{plotmarks}

\definecolor{bluecb}{RGB}{0, 114, 178}
\definecolor{greencb}{RGB}{0, 158, 115}
\definecolor{redcb}{RGB}{204, 121, 167}
\definecolor{orangecb}{RGB}{213, 94, 0}

\begin{tikzpicture}

  \pgfplotsset{
    grid style = {
      dash pattern = on 0.05mm off 1mm,
      line cap = round,
      black,
      line width = 0.5pt
    },
    label style = {font=\fontsize{10pt}{7.2}\selectfont},
    tick label style = {font=\fontsize{9pt}{7.2}\selectfont}
  }

 \begin{semilogyaxis}[%
    xlabel=$E_b/N_0$ (dB),xtick={2,2.5,...,5.0},%
    xlabel style={yshift=0.6em},%
    minor x tick num={1},
    xmin=3,xmax=5,%
    ylabel=FER, ylabel style={yshift=-1.05em},%
    width=0.54\columnwidth, height=6.5cm, grid=major,%
    legend style={
      anchor={center},
      cells={anchor=west},
      column sep=1mm,
      font=\fontsize{8pt}{7.2}\selectfont,
      mark size=3.0pt
    },
    legend columns=5,
    legend to name=qtz-legend,
    mark size=3.0pt, mark options=solid]
    
    \addlegendimage{empty legend}
    \addlegendentry[anchor=east]{\textbf{Floating point:}}

    \addplot[very thick,color=black] table[x=ebn0_db,y=FER] {data/512.427.s0.5.list.L2.C0.float.csv};
    \addlegendentry{}

    \addlegendimage{empty legend}
    \addlegendentry{}

    \addlegendimage{empty legend}
    \addlegendentry{}

    \addlegendimage{empty legend}
    \addlegendentry{}

    \addlegendimage{empty legend}
    \addlegendentry[anchor=east]{\textbf{$Q_i.Q_c.Q_f$:}}

    \addplot[thick,color=bluecb, mark=x] table[x=ebn0_db,y=FER] {data/512.427.s0.5.list.L2.C0.q7.6.1.csv};
    \addlegendentry{$7.6.1$}

    \addplot[thick,color=greencb, mark=o] table[x=ebn0_db,y=FER] {data/512.427.s0.5.list.L2.C0.q6.5.1.csv};
    \addlegendentry{$6.5.1$}

    \addplot[thick,color=redcb, mark=+] table[x=ebn0_db,y=FER] {data/512.427.s0.5.list.L2.C0.q6.5.0.csv};
    \addlegendentry{$6.5.0$}

    \addplot[thick,color=orangecb, mark=diamond] table[x=ebn0_db,y=FER] {data/512.427.s0.5.list.L2.C0.q5.4.0.csv};
    \addlegendentry{$5.4.0$}

  \end{semilogyaxis}
\end{tikzpicture}\hspace{-7pt}
\begin{tikzpicture}
  \pgfplotsset{
    grid style = {
      dash pattern = on 0.05mm off 1mm,
      line cap = round,
      black,
      line width = 0.5pt
    },
    label style = {font=\fontsize{10pt}{7.2}\selectfont},
    tick label style = {font=\fontsize{9pt}{7.2}\selectfont}
  }

 \begin{semilogyaxis}[%
    xlabel=$E_b/N_0$ (dB),xtick={2.0,2.5,...,5.0},%
    xlabel style={yshift=0.6em},%
    minor x tick num={1},
    xmin=3,xmax=5.0,%
    ylabel=BER, ylabel style={yshift=-1.05em},%
    width=0.54\columnwidth, height=6.5cm, grid=major,%
    mark size=3.0pt, mark options=solid]
    
    \addplot[thick,color=black] table[x=ebn0_db,y=BER] {data/512.427.s0.5.list.L2.C0.float.csv};

    \addplot[thick,color=bluecb, mark=x] table[x=ebn0_db,y=BER] {data/512.427.s0.5.list.L2.C0.q7.6.1.csv};

    \addplot[thick,color=greencb, mark=o] table[x=ebn0_db,y=BER] {data/512.427.s0.5.list.L2.C0.q6.5.1.csv};

    \addplot[thick,color=redcb, mark=+] table[x=ebn0_db,y=BER] {data/512.427.s0.5.list.L2.C0.q6.5.0.csv};

    \addplot[thick,color=orangecb, mark=diamond] table[x=ebn0_db,y=BER] {data/512.427.s0.5.list.L2.C0.q5.4.0.csv};

  \end{semilogyaxis}
\end{tikzpicture}
\\
\ref{qtz-legend}
  \caption{Effect of quantization on the error-correction performance of Fast-SSC-List-based decoding of a $(512, 427)$ systematic polar code with list size $L=2$.}
  \label{fig:perf-qtz}
\end{figure}

Fig.~\ref{fig:perf-qtz} shows the effect of quantization on the error-correction performance of List-based decoding of a $(512, 427)$ systematic polar code with list size $L=2$. It can be seen that the coding loss at a \gls{fer} of $10^{-3}$ or a \gls{ber} of $10^{-5}$ can be kept under 0.25~dB with many different configurations. Notably, with $Q_i.Q_c.Q_f=7.6.1$ and $6.5.0$, the coding loss at a \gls{ber} of $10^{-5}$ is approximatively of 0.10~dB and 0.15~dB, respectively. Thus, the proposed implementation uses $Q_i.Q_c.Q_f=6.5.0$ and the path metrics are represented using 7 bits.

\subsection{Comparison with the State of the Art}

Table~\ref{tab:cmp} shows the post-\gls{par} results along with the power consumption estimations for our Fast-SSC-List unrolled decoder implementation. Unfortunately, we could not find implementation results of \gls{scl}-based decoders for a frame length $N=512$ in the literature. Thus, the state-of-the-art works included in Table~\ref{tab:cmp} for comparison are decoders for the closest frame length i.e. $N=1024$. Their list size, however, is identical. It should also be noted that the other works only present synthesis results and their post-\gls{par} results would most likely be slightly worse both in terms of area and in terms of the operating frequency. Since our decoder is for a polar code of higher rate than the other works, we list the coded throughput for fair comparison. Similarly, we also present technology-scaled results, using Dennard scaling laws~\cite{Dennard1974}, for  fair comparison. 

  \begin{table}[t]
    \centering
    \caption{Comparison with state-of-the-art List-based polar decoders. Technology-scaled area results for 28 nm CMOS are included at the bottom.}
    \setlength{\tabcolsep}{0.05cm}
    \resizebox{\columnwidth}{!}{%
      \begin{tabular}{lcccc}
        \toprule
        \textbf{Implementation}&\textbf{this work$^\diamond$}&\phantom{$^\star$}\cite{Yuan2015}$^\star$&\phantom{$^\star$}\cite{Lin2016}$^\star$&\phantom{$^\star$}\cite{Xiong2016}$^\star$\\
        \hline               
        \textbf{Code Length}   		   & 512                & 1,024		& 1,024		& 1,024 \\
        \textbf{Rate} 						   & 0.83               & 0.5 		& 0.5			& 0.5 \\
        \textbf{List Size} 				   & 2                  & 2				& 2			  & 2 \\
        \textbf{Algorithm}	   		   & Approx.            & Exact  	& Approx.	& Approx. \\
        \textbf{Technology}				   & 28 nm              & 65 nm 	& 90 nm		& 90 nm \\
        \textbf{Area} (mm$^2$)		   & 0.87               &	1.06		& 1.98		& 2.32 \\
        \textbf{Supply} (V)          & 1.1                & N/A     & N/A     & N/A \\
        \textbf{Frequency} (MHz)     & 468                & 500     & 423     & 409 \\
        \textbf{Latency} ($\mu$s)    & 0.54               & 2.04    & 0.79    & 0.87\\
        \textbf{Coded T/P} (Gbps)	   & 12.0               & 0.5			& 1.3			& 2.4 \\
        \textbf{Area Eff.} (Gbps/mm$^2$)& 13.79           & 0.47    & 0.67    & 1.04\\
        \textbf{Power} (mW)          & 87                 & 395     & N/A     & N/A\\
        \textbf{Energy Eff.} (pJ/bit)& 7.25               & 790     & N/A     & N/A\\
        \hline               
        \multicolumn{4}{l}{\textit{Normalized results for 28 nm}} &\\
        \hline               
        \textbf{Area} (mm$^2$)       & 0.87               &	0.20 		& 0.19		& 0.22\\
        \textbf{Area Eff.} (Gbps/mm$^2$)& 13.79           & 2.54    & 6.95    & 10.76\\
        \bottomrule
      &&&&\vspace{-8pt}\\
      \multicolumn{4}{l}{\textit{$^\diamond$Post-layout results, 80\% utilization and timing is met.}} &\\
      \multicolumn{4}{l}{\textit{$^\star$Synthesis results.}} &\\
      \end{tabular}
    }
    \label{tab:cmp}
  \end{table}

From Table~\ref{tab:cmp}, it can be seen that the coded throughput of the proposed decoder is from 5 to 21 times higher than that of the other works. Latency is from 32\% to 74\% lower than the other decoders. As the timing constraint was easily met, the clock frequency could be increased to improve throughput and latency at the cost of power consumption. The area of the proposed decoder however is approximately 4 times higher than the normalized area of the $L=2$ List decoders of \cite{Yuan2015,Lin2016,Xiong2016} for $N=1024$. Nonetheless, the post-\gls{par} area efficiency of our decoder is 1.3 to 5.4 times greater than the normalized post-synthesis area-efficiency results of the other works. This efficiency comes at the cost of reduced flexibility: the proposed decoder only supports one specific polar code i.e., the code length or its frozen bit locations cannot be modified at run time. However, the multi-mode idea for unrolled decoders described in \cite{Giard_TCASI_2016} is also applicable to our proposed \gls{scl}-based decoder and support for a few more polar codes could be added. Lastly, the power consumption of our proposed decoder is estimated to be of 87 mW, leading to an energy efficiency of 7.25 pJ/bit.

\section{Conclusion}\label{sec:conclusion}
In this paper, we proposed a List-based decoder hardware implementation for a systematic polar code with better error-correction performance than an LDPC code of similar length and rate from the 802.16e standard.
Post-\gls{par} ASIC results for the 28 nm UTBB-FD-SOI CMOS technology from ST Microelectronics demonstrated that the proposed decoder is capable of sustaining a throughput greater than 10 Gbps with an energy efficiency of 7.25~pJ/bit at a clock frequency of 468~MHz. These results show excellent energy efficiency at the cost of increased area with respect to existing implementations. Yet, the post-\gls{par} area efficiency was shown to be 28\% better than the normalized post-synthesis area efficiency of the best state-of-the-art \gls{scl}-based decoder in the literature.
The key ingredients to achieve these results were to adopt the Fast-SSC-List decoding algorithm to reduce complexity, to adapt the unrolling technique to List-based decoding to increase speed, and to use clock gating to greatly reduce the power consumption.

\balance
\printbibliography

\end{document}